\input stromlo

\title Searching for the Kinematical Signature of a Black Hole in M87

\shorttitle Black Hole Signature

\author David Merritt^1 and Siang Peng Oh^2

\affil @1 Rutgers University 

\affil @2 Princeton University Observatory

\shortauthor Merritt \& Oh

\maintext

Many theories for the formation of massive black holes (BHs) 
in galactic nuclei predict that the stellar motions should 
be anisotropic within the radius of influence of the BH, 
$r_h=GM_h/\sigma_*^2$.
If the BH forms slowly by the accumulation of gas, the 
stellar action variables are adiabatically conserved and the 
orbits become gradually more circular (Young 1980; 
Goodman \& Binney 1984).
The induced anisotropy is quite modest, however, with $\beta = 
1-\sigma_t^2/\sigma_r^2$ typically lying between $-0.3$ and $0$ 
in models where the stellar motions are initially isotropic 
(Quinlan et al. 1995).
In another class of models, nuclear BHs grow through the 
accretion of other BHs acquired from galaxy mergers (e.g. 
Ebisuzaki et al. 1991).
Dynamical friction drags the BHs into the nucleus where they form 
a bound pair.
The binary separation continues to shrink, at first due to 
three-body scattering processes in which the BH binary ejects 
stars, and later through gravitational radiation (Begelman et al. 
1980).
Because stars on elongated orbits are more likely to be ejected, 
three-body scattering introduces a circular bias in the velocity 
distribution of the non-ejected stars (Quinlan 1996a).
Detailed simulations of this process (Quinlan 1996b)
suggest that the induced anisotropy could be much 
greater than in the adiabatic growth model.

Detecting velocity anisotropy in hot stellar systems is generally 
a difficult task, requiring accurate measurements of the 
line-of-sight velocity distribution as a function of position. 
However if the functional form of the gravitational potential is known,
the anisotropy can be inferred from the stellar velocity 
dispersion profile alone (Binney \& Mamon 1982).
In the vicinity of the BH, the potential of a spherical galaxy 
can be approximated as
$$
\Phi(r) = -{GM_h\over r} + \left({M\over L}\right)\Phi_L(r), 
$$
where $M/L$ is the mass-to-light ratio of the stars and $\Phi_L$ 
is the ``potential'' corresponding to the stellar luminosity 
distribution.
Eq. (1) is not completely general, since it assumes that $M/L$ is 
independent of radius.
However it is likely to be accurate within the region where the 
gravitational force is dominated by the BH.
$M/L$ can be determined from the velocity dispersion profile via 
the virial theorem, independent of any assumptions about the velocity 
anisotropy.

Based on observations of a $1''$ ionized gas disk, 
Ford et al. (1994) and Harms et al. (1994) inferred the presence 
of a $2.4 \pm 0.7 \times 10^9 M_{\odot}$ BH in M87.
The radius of influence of this BH would be $r_h\approx 80$ pc $\approx 
1''$; but it would dominate the gravitational {\it force} out to a 
distance of roughly $3 r_h$.
Thus we expect Eq. (1) to be accurate within the region where the 
BH could affect the stellar kinematics.

Fig. 1 shows estimates of $\beta(r)$ for the stars 
in M87 based on van der Marel's (1994) velocity dispersion measurements.
The data were corrected for seeing and for instrumental blurring 
using a regularized algorithm (Merritt \& Oh 1996).
For $M_h\geqsim 1.0\times 10^9 M_{\odot}$, the stellar motions 
are significantly anisotropic, $\sigma_t\geqsim\sigma_r$, within 
$1''-2''$.
The anisotropy is greater than predicted by the adiabatic growth 
models but may be consistent with the predictions of the binary BH 
model.

\figureps[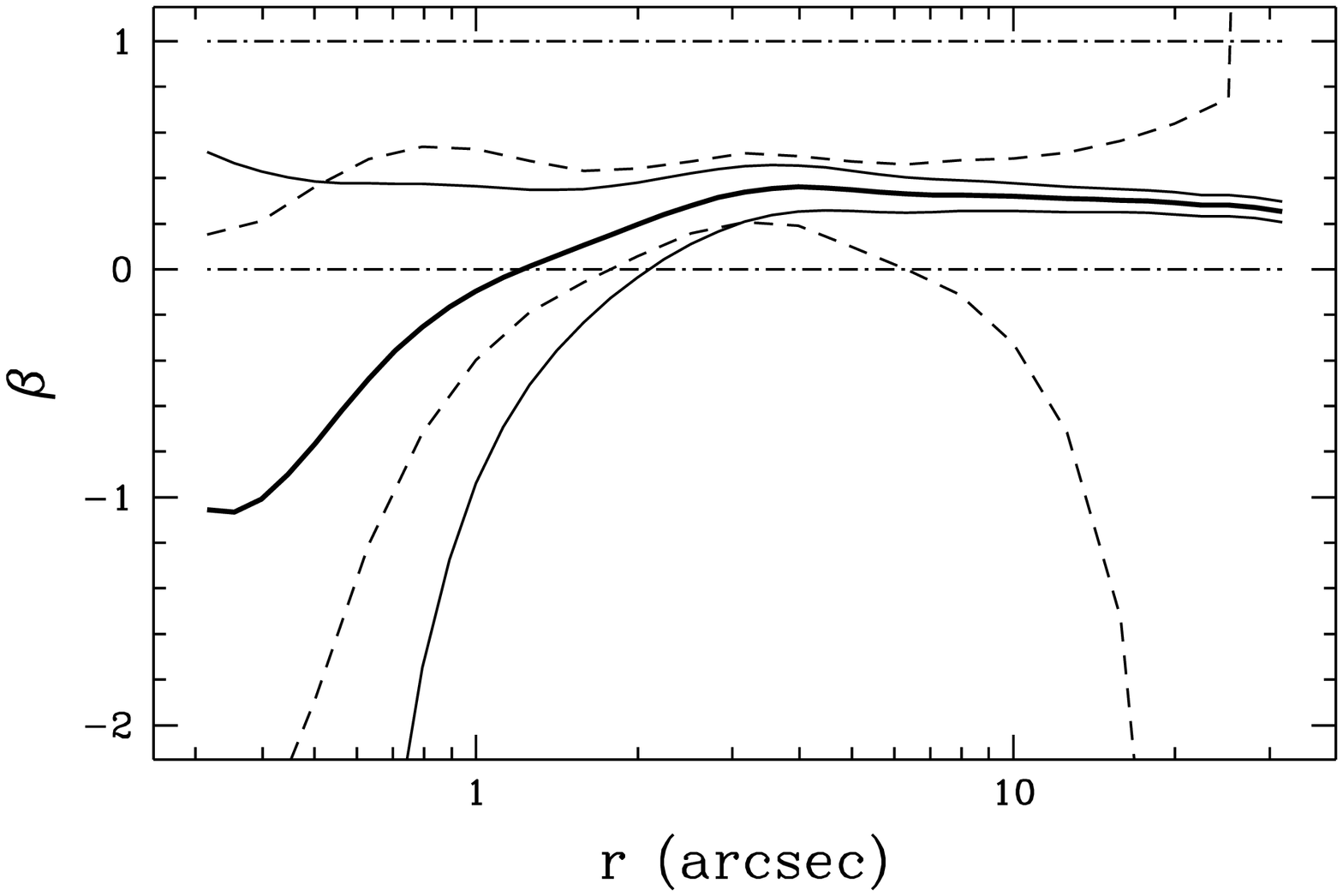,.4\vsize] 1. Anisotropy parameter 
$\beta = 1-\sigma_t^2/\sigma_r^2$ as a function of radius for the
stellar motions in M87.
Curves are for assumed black hole masses $M_h/M_{\odot}=1.0\times 10^9$ 
(upper), $2.4\times 10^9$ (heavy, with 95\% confidence bands) 
and $3.8\times 10^9$ (lower). 
The velocity dispersion data from which these curves were derived
extends only to $25''$; hence, the confidence bands become very wide
at large radii.

\references

Begelman, M. C., Blandford, R. D. \& Rees, M. J. 1980, Nature, 287, 307

Binney, J. J. \& Mamon, G. A. 1982, MNRAS, 200, 361

Ebisuzaki, T., Makino J., \& Okumura, S. K. 1991, Nature, 354 212

Ford, H. C., et al. 1994, ApJ, 435, L27

Goodman, J. \& Binney, J. J. 1984, MNRAS, 207, 511

Harms, R. J., et al. 1994, ApJ, 435, L35

Merritt, D. \& Oh, S.-P., submitted to AJ

Quinlan, G. D. 1996a, New Astronomy, 1, 35

Quinlan, G. D. 1996b, in preparation

Quinlan, G. D., Hernquist, L. \& Sigurdsson, S. 1995, ApJ, 440, 554

van der Marel, R. P. 1994, MNRAS, 270, 271

Young, P. 1980, ApJ, 242, 1232

\bye